\documentclass{article}
\newcommand{\be}{\begin{equation}}    
 \newcommand{\ee}{\end{equation}}
\def\Hscr{{\cal H}} \def\Nscr{{\cal N}} \def\HNscr{\Hscr_\Nscr}
\def\IN{I_\Nscr}
\newcommand{\sizedef}{
        \headheight=0pt                               
 	  \topmargin=-1.5cm \headsep=1.5cm              
        \oddsidemargin=-0.5cm \evensidemargin=-0.5cm  
        \textheight=22truecm \textwidth=16.5truecm    
 	  \setlength{\columnsep}{20pt}                  
  }

\catcode`@=11

\newcommand\eqalign[1]{\null\,\vcenter{\openup\jot\m@th
  \ialign{\strut\hfil$\displaystyle{##}$&$\displaystyle{{}##}$\hfil
      \crcr#1\crcr}}\,}
      \newcommand\meqalign[1]{\null\,\vcenter{\openup\jot\m@th
  \ialign{\strut\hfil$\displaystyle{##}$&&$\displaystyle{{}##}$\hfil
      \crcr#1\crcr}}\,}
      \newcommand\case[2]{\textstyle{\frac{#1}{#2}}}
      \newcommand\sgn{\mathop{\rm sgn}\nolimits}
\def\twoR{\,{}^{(2)}\negthinspace R}

\catcode`@=12   

\newdimen\arrayruleHwidth
\makeatletter
\newcommand\Hline{\noalign{\ifnum0=`}\fi\hrule \@height \arrayruleHwidth
  \futurelet \@tempa\@xhline}
\makeatother
\sizedef

\begin{document}
\bibliographystyle{prsty}


\title{Invariants at fixed and arbitrary energy.\\
       A unified geometric approach.}
\author{Kjell Rosquist\thanks{Supported by the
        Swedish Natural Science Research Council. E-mail: kr@physto.se.} \\
        Department of Physics, Stockholm University, \\
        Albanova University Center, Stockholm, Sweden \\
         and \\
        Giuseppe Pucacco\thanks{Supported by INFN (Istituto Nazionale di Fisica Nucleare). E-mail: pucacco@roma2.infn.it.}  \\
        Physics Department, University of Rome "Tor Vergata",\\
        Via della Ricerca Scientifica, 1,
        I-00133 Rome, Italy}
        \date{}
\maketitle

\vskip2cm

Published as: {\it J. Phys. A: Math. Gen. {\bf 28}, 3235--3252  (1995)} 

\vskip2cm

\begin{abstract}

\noindent
Invariants at arbitrary and fixed energy (strongly and weakly conserved
quantities) for 2-dimensional Hamiltonian systems are treated in a unified
way. This is achieved by utilizing the Jacobi metric geometrization of the
dynamics. Using Killing tensors we obtain an integrability condition for
quadratic invariants which involves an arbitrary analytic function $S(z)$.
For invariants at arbitrary energy the function $S(z)$ is a second degree
polynomial with real second derivative. The integrability condition then
reduces to Darboux's condition for quadratic invariants at arbitrary energy.
The four types of classical quadratic invariants for positive definite
2-dimensional Hamiltonians are shown to correspond to certain conformal
transformations. We derive the explicit relation between invariants in the
physical and Jacobi time gauges. In this way knowledge about the invariant in
the physical time gauge enables one to directly write down the components of
the corresponding Killing tensor for the Jacobi metric. We also discuss the
possibility of searching for linear and quadratic invariants at fixed energy
and its connection to the problem of the third integral in galactic
dynamics. In our approach linear and quadratic invariants at fixed energy can
be found by solving a linear ordinary differential equation of the first or
second degree respectively.

\end{abstract}


\clearpage

\section{Introduction}

The quest for integrable systems is still one of the main areas of interest
in  classical dynamics both per se and for the applications in the related
fields of celestial mechanics, accelerator physics and so on. In galactic
dynamics a long-standing and to this day unresolved issue concerns the
celebrated third integral. Briefly, the observed motion of stars in the galaxy
has a certain regularity which indicates the existence of a third invariant
in addition to the energy and angular momentum. This picture is
confirmed by numerical simulations. In modern terminology, the motion
is non-chaotic contrary to what one would expect for a generic potential. A
particularly striking numerical experiment was made by H\'enon and Heiles
\cite{hh:inv} in 1964 using a model potential now known as the
H\'enon-Heiles potential. They showed that the motion in that
potential is regular with invariant tori up to a certain energy value
$E_0$. At this energy the invariant tori abrubtly begin to break up and the
motion becomes chaotic in increasingly large parts of the phase space. One
conclusion is that there can be no invariant which commutes with the
Hamiltonian. On the other hand the very high degree of regularity at low
energies, $E < E_0$, seems to indicate the existence of an additional
invariant at those energies. Such an invariant, if it exists, must
necessarily depend on the energy with some kind of singular behaviour at $E
= E_0$. However, no such invariant is known, for any value of the energy.
One of the purposes of this paper is to discuss the possibility to find
invariants at fixed energy values.

An essential aspect of integrability is therefore to check for the existence
of integrals of the  motion in addition to the energy, namely phase-space
functions invariant with  respect to the phase flow. The search for additional
invariants has therefore been actively pursued for more than a century
(Bertrand, 1852
\cite{bertrand:inv}; Darboux, 1901 \cite{darboux:inv}; for  a complete review
of the matter see Hietarinta \cite{hietarinta:secinv}). A less known aspect
of the question is the relation between the invariants that most frequently
appear in the applications, namely polynomials in the momenta, and Killing
tensors of the configurational manifold of the dynamical system. This
relation is in the form of a straightforward correspondence if the Jacobi
approach to the geometrization of the dynamics of a conservative system is
adopted.

The framework of the Jacobi geometrical formulation of the dynamics offers
several conceptual and technical aids that give the possibility of shedding
new light on a number of aspects of integrability. From the technical point of
view, it  allows the use of the powerful tools of Riemannian geometry on the
configurational manifold of the corresponding natural Jacobi formulation of
the dynamical system. From the conceptual point of view, it treats invariants
at fixed energy (the so called configurational invariants of Hall
\cite{hall:inv}) on the same footing as the invariants at arbitrary  energy,
thus realizing a unified geometric approach.

One aim of the present paper is to demonstrate the fruitfulness of the
geometric approach by providing the generalization of Darboux's conditions to
include the configurational invariants. In the particular but fundamental
case of two degrees of freedom, the Killing equations for second rank
Killing tensors can be solved in full generality, resulting in all the known
cases of integrability at arbitrary energy, both the classical ones already
obtained by Darboux \cite{darboux:inv} and those recently found to exhaust
all the possibilities in two dimensions (Dorizzi et al.\ \cite{dgr:inv}).

The substantially larger family of solutions of the Killing equation in the
fixed energy case allows one to find large new classes of constrained
integrable  systems, with a backwards procedure going from the knowledge of
the form of the invariants to the assessment of the structure of the
potentials admitting  constrained integrability. Applying a new prescription
for transforming Hamiltonian symmetries between different time gauges, the
relation between the invariants in the physical and Jacobi time gauges is
explicitly determined. The relation between the invariants also provides a
shortcut to calculate the Killing tensor from the knowledge of the invariant
in the physical time gauge.

\section{Geometric representation of the dynamics}
\label{sec:geom}

The geometrization of the dynamics in terms of the Jacobi geometry
has been known for about a century and can be found in some text books
\cite{am:mechanics,arnold:mechanics,lanczos:mechanics}. In spite of
this it is not widely known or used and therefore we outline the main ideas
in this section to make the paper self-contained. Most Hamiltonian systems of
physical interest have a geometric kinetic energy part, that is to say that
the kinetic energy is a non-degenerate quadratic form in the momenta
$p_\alpha$
\be
     T = \frac12 h^{\alpha\beta} p_\alpha p_\beta \ ,
\ee
where $h_{\alpha\beta}$ is a function of the configuration variables
$q^\alpha$. The Hamiltonian itself then has the form
\be\label{eq:origham}
     H = T + V(q) = E \ .
\ee
The independent variable $t$ is often but not always the time. For
simplicity we shall refer to the independent variable as the time in this
paper. For any given energy $E$ of the system we can use the Hamiltonian
\be
     \Hscr = H-E \ ,
\ee
to represent the dynamics provided that we impose the constraint
\be
     \Hscr = 0 \ .
\ee
For any such zero energy Hamiltonian we can reparametrize the system by
introducing a new time variable $t_\Nscr$ defined by the relation
\be
     dt = \Nscr(p,q) dt_\Nscr \ ,
\ee
together with a redefined Hamiltonian
\be
       \HNscr = \Nscr(p,q) \Hscr = \Nscr T + \Nscr (V-E) = 0 \ .
\ee
The new Hamiltonian will then give the same equations of motion on the
constraint surface $\HNscr=0$ (see e.g.\ \cite{urj:geom}). We shall
use the term {\it lapse function\/} for $\Nscr(p,q)$ which defines the
independent variable gauge. This usage is borrowed from  applications in
general relativity where the lapse gives the rate of physical time change
relative to coordinate time (see e.g.\ \cite{mtw:gravitation}). The
lapse function can be taken as any non-zero function on the phase space.

The dynamics of the system can be represented in a purely geometric
formulation by exploiting the reparametrization freedom. The passage to the
geometric representation is accomplished by defining a new time variable
$t_J$ ({\it Jacobi time}) by the lapse choice
\be
    \Nscr = \Nscr_J = [2(E-V)]^{-1} \ .
\ee
Note that for a positive definite kinetic energy $E-V$ is always nonnegative
in the physically allowed region. The corresponding Hamiltonian has the form
$\HNscr = [2(E-V)]^{-1} T - \frac12$. Thus $\HNscr$ has a constant potential
energy  which can be subtracted without affecting the equations of motion.
This leads to the {\it Jacobi Hamiltonian\/}, $H_J$, defined by
\be
     H_J = \frac12 \Nscr_J h^{\alpha\beta} p_\alpha p_\beta
         = \frac12 J^{\alpha\beta} p_\alpha p_\beta \ ,
\ee
where $J^{\alpha\beta}$ is the {\it Jacobi metric\/} of the system. This
defines a geometry which contains all information about the dynamics. The
dynamical metric is given in covariant form by
\be\label{eq:jacobimetric}
     J_{\alpha\beta} = 2(E-V) h_{\alpha\beta} \ .
\ee
The Jacobi metric is therefore conformally related to the original kinetic
metric $h_{\alpha\beta}$.  It follows that the dynamics of the system
(\ref{eq:origham}) is equivalent to geodesic motion in the Jacobi
geometry\footnote { It may happen that there are other geometries which can
also be used to represent the dynamics. In that case the dynamical geometry
defined according to the prescription given here is not unique. For example
an inequivalent dynamical geometry may sometimes be obtained by performing a
suitable canonical transformation in some time gauge (see e.g.\
\cite{atu:miniquant}).}
\be
     ds^2 = J_{\alpha\beta} dq^\alpha dq^\beta \ .
\ee
Note that the Jacobi geometry depends on the energy parameter $E$. This means
that in general the Jacobi geometries corresponding to different energy
surfaces are inequivalent. The complete geometric representation of the
dynamics of the system is therefore given by the geodesics of a 1-parameter
family of geometries. Because of the form of the conformal factor in
the Jacobi geometry (\ref{eq:jacobimetric}) the geometric representation is
only valid locally in configuration space at points where $V\neq E$. In
general the equation $V=E$ defines a non-empty energy surface in
configuration space for a given value of $E$. Exceptions occur if e.g.\ $V>0$
throughout the configuration space and $E\leq 0$. In this paper we are only
concernced with the local existence of invariants. Therefore the failure of
the Jacobi geometry to represent the global dynamics does not affect our
analysis.

\section{Positive definite two-dimensional Hamiltonian systems}
\label{sec:posdef}

To study 2-dimensional systems it is very helpful to use variables
which are null (lightlike) with respect to the dynamical metric. Such
variables are naturally adapted to the action of the conformal group
which plays an essential role for 2-dimensional systems. The indefinite
(Lorentzian) signature case was discussed by Rosquist and Uggla
\cite{ru:kt}. The null variables for that case are real and the conformal
group can be parametrized by two arbitrary real functions of one variable.
Our approach is to use the method of \cite{ru:kt} to treat the case
of a positive definite dynamical metric.

We consider a general 2-dimensional positive definite dynamical
metric written in the manifestly conformally flat form
\be
     ds^2 = 2G(x,y) (dx^2 + dy^2) \ .
\ee
We seek a condition on $G$ which guarantees the existence of a second rank
Killing tensor. To that end we follow \cite{ru:kt} as closely as possible and
introduce null variables which in the positive definite case are complex
\be\label{eq:complex}\eqalign{
           z &= x + iy \ ,\cr
         \bar z &= x - iy \ .\cr
}\ee
Here and in the following a bar is used to denote complex conjugation. The
metric then becomes
\be\label{eq:ds2complex}
    ds^2 = 2G(z,\bar z)dz d\bar z \ .
\ee
Note that although our notation is similar to that in \cite{ru:kt}
the variables $z$ and $\bar z$ are complex in the positive definite case as
opposed to the Lorentzian case where they are real.

When doing calculations it is convenient to employ either an orthonormal frame
$\omega^{\hat I}$ using hatted upper case Latin indices ($\hat I,\hat J,\hat
K,\ldots = \hat1, \hat2$) in terms of which the metric is written as
\be  ds^2 = (\omega^{\hat1})^2 + (\omega^{\hat2})^2 \ ,
\ee
or a complex null frame $\Omega^I$ using upper case Latin indices
($I,J,K,\ldots = 1,2$) with the metric written as
\be  ds^2 = 2 \Omega^1 \Omega^2 \ .
\ee
The frame components are given by
\be\label{eq:frame}\meqalign{
     \omega^{\hat1} &= (2G)^{1/2}dx\ ,\qquad \Omega^1 &= G^{1/2}dz\ ,\cr
     \omega^{\hat2} &= (2G)^{1/2}dy\ ,\qquad \Omega^2 &= G^{1/2}d\bar z\ ,\cr
}\ee
The two frames are related by $\Omega^I = e^I{}_{\hat I} \omega^{\hat I}$ or
$\omega^{\hat I} = e_I{}^{\hat I} \Omega^I$ where $e_I{}^{\hat I}$ is the
transposed matrix inverse of $e^I{}_{\hat I}$. The transformation matrix is
given by
\be  (e^I{}_{\hat I}) = \left(\matrix{
                           \frac1{\sqrt2}&  \frac{i}{\sqrt2}& \cr
                           \frac1{\sqrt2}& -\frac{i}{\sqrt2}& \cr} \right)
                           \ ,\qquad
         (e_I{}^{\hat I}) = \left(\matrix{
                           \frac1{\sqrt2}& -\frac{i}{\sqrt2}& \cr
                           \frac1{\sqrt2}&  \frac{i}{\sqrt2}& \cr} \right)\ .
\ee

The simplest symmetry of the geodesic equations is known as a Killing vector
field (see e.g.\ \cite{schutz:geom}). A second rank Killing tensor is a
symmetric tensor $K_{IJ} = K_{(IJ)}$ satisfying the equation
\be
     K_{(IJ;K)} = 0 \ .
\ee
The existence of such a tensor is equivalent to the existence of a polynomial
invariant of the second degree $K^{IJ} p_I p_J$ for the geodesic equations.
Killing tensors and second degree invariants are the natural generalizations
of Killing vectors $K^I$ and the corresponding first degree invariants $K^I
p_I$. Any Killing vector gives rise to a Killing tensor $K_{(I} K_{J)}$. Such
a Killing tensor is said to be reducible. An important property of the second
rank Killing tensor equations is that they can be decomposed in conformal
(traceless) and trace parts according to
\be\label{eq:kteq}\eqalign{
     & P_{(IJ;K)} - \frac12 h_{(IJ} P^L{}_{K);L} = 0 \ ,\cr
     & K_{;I} = - P^J{}_{I;J} \ ,\cr
}\ee
where the Killing tensor itself is decomposed in a conformal part $P_{IJ}$ and
the trace $K = K^I{}_I$ according to
\be
     K_{IJ} = P_{IJ} + \frac12 K h_{IJ} \ .
\ee
Referring to the vector $P_I  = P^J{}_{I;J}$ as the conformal current it
follows from (\ref{eq:kteq}) that for any given conformal Killing tensor the
equation for the trace can be solved if the integrability condition
\be\label{eq:intcond1}
     P_{[I;J]} = 0
\ee
is satisfied. The procedure to solve the Killing tensor equations is
therefore to first solve the conformal Killing tensor equations and then
check if the integrability condition (\ref{eq:intcond1}) for the trace can be
satisfied. In fact it turns out that the conformal equations can easily be
solved leaving the integrability condition as the remaining equation to study.

Let $K_{\hat I \hat J}$ be the components of a Killing tensor with respect to
the orthonormal frame $\omega^{\hat I}$. Then the null frame components are
given by $K_{IJ} = e_I{}^{\hat M} e_J{}^{\hat N} K_{\hat M \hat N}$ which
gives
\be\eqalign{
     K_{11} &= \frac12(K_{\hat1\hat1}-K_{\hat2\hat2})-iK_{\hat1\hat2}\ ,\cr
     K_{22} &= \frac12(K_{\hat1\hat1}-K_{\hat2\hat2})+iK_{\hat1\hat2}\ ,\cr
     K_{12} &= K/2 = \frac12(K_{\hat1\hat1} + K_{\hat2\hat2}) \ ,\cr
   }\ee
where $K = K^I{}_I$ is the trace. Since the orthonormal components $K_{\hat
I \hat J}$ are by definition real it follows that $K_{22} =
\overline{K_{11}}$. Therefore we can represent the null frame components of
the conformal part of the Killing tensor by a single complex function
$S(z,\bar z)$ according to (cf.\ \cite{ru:kt})
\be\label{eq:kt}
       (K_{MN}) = \left( \matrix{\bar S G &  K/2& \cr
                                      K/2 &  S G& \cr} \right)\ ,
\ee
the trace $K$ being a real function. The conformal Killing tensor equations
are formally identical to those obtained in the Lorentzian case \cite{ru:kt}
\be\label{eq:confeq}
         C_{111} = G^{1/2} \bar S_{,z} = 0 \ , \qquad
         C_{222} = G^{1/2} S_{,\bar z} = 0 \ .
\ee
It follows that $P_{MN}$ is a conformal Killing tensor precisely if $S$ is
a function of $z$ only. One way of stating this result is that the equations
(\ref{eq:confeq}) for the conformal Killing tensor coincide with
Cauchy-Riemann's equations for certain rescaled linear combinations of the
tracefree parts of the Killing tensor. This result was implicit in
\cite{kolokoltsov:inv}. Thus the conformal Killing tensor has the form
\be
       (P_{MN}) = \left( \matrix{\bar S(\bar z) G &       0 & \cr
                                                0 &  S(z) G & \cr} \right)\ .
\ee
The trace equations can then be written in the form
\be\label{eq:traceeq}\eqalign{
      K_{,z} &= - 2 \bar S(\bar z) G_{,\bar z} - G \bar S'(\bar z) \ ,\cr
 K_{,\bar z} &= - 2 S(z) G_{,z} - G S'(z) \ ,\cr
}\ee
the derivative $K_{,\bar z}$ being given by taking the complex conjugate of
the above equation.

As in the Lorentzian case there is an infinite-dimensional family of
conformal Killing tensors in this case parametrized by the single complex
analytic function $S(z)$. Following the procedure in \cite{ru:kt} we now write
down the integrability condition for the conformal current. This is exactly
what we need to guarantee the existence of a Killing tensor. In the positive
definite case the integrability condition becomes
\be\label{eq:intcond2}\eqalign{
     P_{[1;2]} = &- G^{-1}G_{,zz} S(z)
                 + G^{-1} G_{,\bar z \bar z} \bar S(\bar z)
                 - \frac32 G^{-1} G_{,z} S'(z)
                 + \frac32 G^{-1} G_{,\bar z} \bar S'(\bar z) \cr
                 &- \frac12 S''(z) + \frac12 \bar S''(\bar z)
               =   0 \ ,\cr
}\ee
or in a more compact form
\be  P_{[1;2]} = -i\, \Im \{ 2G^{-1}G_{,zz}S(z) + 3G^{-1}G_{,z}S'(z)
                               + S''(z) \} = 0 \ ,
\ee
where $\Im$ denotes the imaginary part. As in \cite{ru:kt} we use
a conformal transformation to standardize the frame and coordinate
representation of the conformal Killing tensor. To that end we introduce the
new null frame $\tilde\Omega^1 = B\Omega^1$, $\tilde\Omega^2 = B^{-1}\Omega^2$
and a new complex coordinate by the transformation $w=H(z)$ with inverse
$z=F(w)$. By choosing $B=(\bar F'(\bar w)/ F'(w))^{1/2}$ we ensure that the
new frame $\tilde\Omega^I$ has the coordinate representation $\tilde\Omega^1
= \tilde G^{1/2} dw$,  $\tilde\Omega^2 = \tilde G^{1/2} d\bar w$ where
$\tilde G = |F'(w)|^2 G$ is the new metric conformal factor, $ds^2 = 2\tilde G
dw d\bar w$. It follows that the conformal Killing tensor components in the
new frame are given by
\be\eqalign{
        \tilde P_{11} &= B^{-2} P_{11}
                       = [\bar H'(\bar z)]^2 \bar S(\bar z) \tilde G\ ,\cr
        \tilde P_{22} &= B^2 P_{22} = [H'(z)]^2 S(z) \tilde G\ .\cr
   }\ee
The coordinates are standardized by choosing a conformal transformation
function $H(z)$ which satisfies $[H'(z)]^{-2} = S(z)$ implying that $\tilde
P_{11} = \tilde P_{22} = \tilde G$. Note that $\det(P_{IJ})$ is always
positive. Therefore there is only one type of Killing tensors for positive
definite Hamiltonians unlike the situation in the Lorentzian case (see
\cite{ru:kt}). For a standardized conformal Killing tensor the integrability
(\ref{eq:intcond2}) condition simplifies to
\be  \tilde G_{,ww} = \tilde G_{,\bar w \bar w} \ .
\ee
Writing $w = X + iY$ the solution of this equation is
\be  \tilde G = Q_1(X) + Q_2(Y) \ .
\ee
This is the usual form of the potential in separable coordinates.

The trace equation [see (\ref{eq:kteq})] can then be written
\be\eqalign{
       K_{,X} &= - 2 \tilde G_{,X} \ ,\cr
       K_{,Y} &=   2 \tilde G_{,Y} \ ,\cr
   }\ee
with the solution
\be  K = -2 Q_1(X) + 2 Q_2(Y).
\ee
The integrability condition (\ref{eq:intcond2}) can be interpreted as follows.
Given any analytic function $S(z)$ there is a family of potentials (the
solutions of (\ref{eq:intcond2})) which is integrable at zero energy. For
future reference we note that the original potential is given in terms of the
standardized potential by the relation
\be\label{eq:potrel}
     G = |S(z)|^{-1} \tilde G \ .
\ee

We wish to express the invariant $I_J = K^{MN} p_M p_N$ in terms of the
coordinate momentum components. The subscript $J$ is used to distinguish
the invariant in the Jacobi time gauge from the invariant $I$ in the
physical time gauge (see section \ref{sec:timetrans}). By (\ref{eq:frame}) the
null frame components are related to the complex coordinate components by
$(p_1, p_2) = G^{-1/2} (p_z, p_{\bar z})$. The invariant can then be written
in complex coordinates as
\be
     I_J = S(z) p_z{}^2 + \bar S(\bar z) p_{\bar z}{}^2
                      + G^{-1} K p_z p_{\bar z} \ .
\ee
Using the relation
\be\eqalign{
           p_z &= \frac12 (p_x - i p_y) \ ,\cr
    p_{\bar z} &= \frac12 (p_x + i p_y) \ ,\cr
}\ee
implied by (\ref{eq:complex}) we obtain the invariant in real coordinates as
\be\label{eq:invreal}
     I_J = \frac12 \Re(S) (p_x{}^2 - p_y{}^2) + \Im(S) p_x p_y
                                 + \frac14 G^{-1} K (p_x{}^2 + p_y{}^2) \ .
\ee
Note that the last term is equal to $K H_J$.

\section{Arbitrary energy invariants}
A common situation is that one is interested in invariants which are
valid for arbitrary values of the energy. Invariants of that type
arise if the integrability condition itself is independent of the energy. This
happens precisely if $\Im\{S''(z)\} = 0$. Since $S''(z)$ is an analytic
function it follows that $S''(z)$ is a real constant and therefore $S(z)$
must be a second degree polynomial
\be\label{eq:poly}
     S(z) = W(z) := a z^2 + \beta z + \gamma \ ,
\ee
where $a$ is a real constant and $\beta$ and $\gamma$ are complex constants.
The integrability condition (\ref{eq:intcond2}) then simplifies to
\be\label{eq:darboux}
    \Im\{ 2G_{,zz} S(z) + 3G_{,z} S'(z) \} = 0 \ .
\ee
This is nothing but Darboux's condition for quadratic constants of the
motion \cite{darboux:inv,hietarinta:secinv} written in complex variables. The
general integrability condition (\ref{eq:intcond2}) therefore generalizes
Darboux's condition to include also quadratic invariants at fixed energy.

For a given analytic function $S(z)$ the potential may be written as
$G(X,Y) = |F'(w)|^{-2} (Q_1(X) + Q_2(Y))$. In the arbitrary energy case we
have $S(z) = W(z) = [H'(z)]^{-2}$. The equation $H'(z) = [W(z)]^{-1/2}$ can
then be integrated resulting in (without loss of generality we may put $a=1$
if $a\neq0$ since (\ref{eq:darboux}) is invariant under a real scaling of
$S$)
\be\label{eq:hz}
     w = H(z) = \cases{
        \log(\sqrt{W} + z + \beta/2) + const \ , & $(a=1)$ \ ,\cr
        2\beta^{-1}\sqrt{\beta z + \gamma} + const
                                   \ , & $(a=0,\beta\neq0)$ \ ,\cr
        \gamma^{-1/2} z +const     \ , & $(a=0,\beta=0)$
\ .\cr }\ee
Inverting these relations we must distinguish between the four cases
\be\eqalign{
      \hbox{(i)}   \quad & a=1, \Delta:=\sqrt{\beta^2/4-a\gamma}\neq0 \ ,\cr
      \hbox{(ii)}  \quad & a=1, \Delta=0 \ ,\cr
      \hbox{(iii)} \quad & a=0,\beta\neq0  \ ,\cr
      \hbox{(iv)}  \quad & a=0,\beta=0 \ ,\cr
}\ee
giving rise to the following explicit formulas for $F(w)$
\be  z = F(w) = \cases{
        \Delta \cosh(w-w_0)-\beta/2 \ ,             & (i)   \cr
        e^{w-w_0} - \frac{\beta}2 \ ,               & (ii)  \cr
        \frac14 \beta (w-w_0)^2 -\gamma/\beta  \ ,  & (iii) \cr
        \gamma^{1/2} w-w_0 \ .                      & (iv)  \cr
}\ee
where $w_0$ is an arbitrary constant. We are primarily interested in proper
conformal transformation, i.e. $S(z) \neq 1$. Therefore we are free to
perform translations and rotations in $z$ and $w$ to simplify formulas. In
particular we may put $w_0$ equal to zero by a translation of the origin in
the $w$-plane. We can also rotate the $z$-plane to transform $\Delta$ to a
positive real number in case (i). Likewise $\beta$ may be assumed real in
case (iii) and $\gamma$ can be taken as real and positive in case (iv). The
argument we used to set $a=1$ can then be used to set $\beta=4$ and
$\gamma=1$ for convenience in cases (iii) and (iv) respectively. We also
translate the $z$-origin obtaining finally
\be\label{eq:fw}
     F(w) = \cases{
               \Delta \cosh w  \ , & (i)   \cr
               e^w             \ , & (ii)  \cr
               w^2             \ , & (iii) \cr
               w               \ , & (iv)  \cr}
\ee
where $\Delta$ is now to be understood as a positive real number.
Differentiation of the expressions (\ref{eq:fw}) yields
\be
         F'(w) = \cases{ \Delta \sinh w \ , & (i)   \cr
                         e^w            \ , & (ii)  \cr
                         2w             \ , & (iii) \cr
                         1              \ , & (iv)  \cr
}\ee
implying that the conformal transformation factors take the forms
\be
    |S(z)| = |F'(w)|^2 =
           \cases{\Delta^2(\sinh^2 X + \sin^2 Y)  =
                \sqrt{(r^2+\Delta^2)^2-4\Delta^2 x^2} \ , & (i)   \cr
           e^{2X} = r^2                               \ , & (ii)  \cr
           4(X^2+Y^2) = 4r                                \ , & (iii) \cr
           1                                          \ , & (iv)  \cr
}\label{eq:fpw}\ee
where $r := \sqrt{x^2+y^2}$.

In case (i) the conformal coordinate transformation is given by
\be\label{eq:stackelconf}\eqalign{
         x &= \Delta \cosh X \cos Y \ ,\cr
         y &= \Delta \sinh X \sin Y \ .\cr
}\ee
This is the transformation used to define the St\"ackel potential
\cite{bt:galdyn} which was also discussed by Darboux \cite{darboux:inv}. Case
(i) therefore gives St\"ackel's classical integrable potential. From
(\ref{eq:fpw}) and (\ref{eq:potrel}) we find that the potential is given by
the well-known formula \cite{hietarinta:secinv}
\be
     G(x,y) = \frac{Q_1(X(x,y)) + Q_2(Y(x,y))}
                       {\Delta^2[\sinh^2 X(x,y) + \sin^2 Y(x,y)]}
            = \frac{A(\xi(x,y)) + B(\eta(x,y))}{\xi(x,y)+\eta(x,y)} \ ,
\ee
where
\be\eqalign{
     \xi(x,y)  &:= 2\Delta^2 \sinh^2 X = r^2 - \Delta^2
             + \sqrt{(r^2+\Delta^2)^2-4\Delta^2 x^2} \ ,\cr
     \eta(x,y) &:= 2\Delta^2 \sin^2 Y = -r^2 + \Delta^2
             + \sqrt{(r^2+\Delta^2)^2-4\Delta^2 x^2} \ .\cr
}\ee
and the functions $A$ and $B$ are arbitrary functions of their arguments. The
case (ii) conformal transformation is given by
\be\eqalign{
      x &= e^X \cos Y \ ,\cr
      y &= e^X \sin Y \ .\cr
}\ee
In particular it follows that $Y = \arctan(y/x) =: \phi$ is a polar angle.
Using (\ref{eq:fpw}) it then follows that the potential can be written as
\be
     G(x,y) = A(r) + r^{-2} B(\phi) \ ,
\ee
where $A$ and $B$ are arbitrary functions. This is again a classical
integrable case \cite{hietarinta:secinv} known as the Eddington potential.
Case (iii) is characterized by the conformal transformation
\be\eqalign{
      x &= X^2 - Y^2 \ ,\cr
      y &= 2XY       \ .\cr
}\ee
Using (\ref{eq:fpw}) this gives immediately the likewise well-known
classical integrable potential \cite{hietarinta:secinv}
\be
     G(x,y) = [A(r+x) + B(r-x)]/r \ ,
\ee
where again $A$ and $B$ are arbitrary functions. Finally in case (iv) the
conformal transformation factor is unity and so the potential is simply given
by the explicitly separated form
\be
     G(x,y) = A(x) + B(y) \ ,
\ee
in terms of the arbitrary functions $A$ and $B$. This completes the list of
the four classical cases. The three nontrivial conformal transformations
giving rise to systems with a second linear or quadratic invariant are
tabulated in table \ref{tab:conf}.

\begin{table}[tbp]
\begin{center}
\begin{tabular}{cccc}
	\Hline
	 case & \begin{tabular}{c}
         conformal transformation \\
         $w = H(z)$ \end{tabular}         & \begin{tabular}{c}
                                        conformal Killing tensor component \\
                                         $S(z) = [H'(z)]^{-2}$ \end{tabular}
                                                    & Separating coordinates \\
 \hline
	(i)   & $\ln{z \pm \sqrt{z^2-\Delta^2}}$ & $z^2-\Delta^2$ & Elliptical \\
	(ii)  & $\ln z$                          & $z^2$          & Spherical \\
	(iii) & $\sqrt{z}$                       & $4z$           & Parabolic         \\
	\Hline
\end{tabular}
	\caption{\protect\small\protect\parbox[t]{12cm}{
          Conformal transformation functions giving rise to systems with a
          second invariant which is linear or quadratic in the momenta.}}
	\protect\label{tab:conf}
\end{center}
\end{table}

\section{The Killing vector subcase}
Although Killing vectors correspond to reducible second rank Killing tensors
it is nevertheless worthwhile to give a separate treatment of that subcase.
As will be shown below the integrability condition for Killing vectors is a
first order differential equation. In some applications it can
advantageous to investigate this simpler case before tackling the second
rank Killing tensors. A Killing vector can be described by a function
$Z(z,\bar z)$ according to (cf.\ \cite{ru:kt})
\be\eqalign{
              K_1 &= G^{1/2} \bar Z \ ,\cr
              K_2 &= G^{1/2}      Z \ .\cr
}\ee
The Killing vector equations then become
\be
      \overline{K_{(1;1)}} = K_{(2;2)} = Z_{,\bar z} = 0 \ ,
\ee
\be\label{eq:kveq}
          K_{(1;2)} = \frac12 \left( Z_{,z} + \bar Z_{,\bar z}
            + G^{-1} G_{,z} Z + G^{-1} G_{,\bar z} \bar Z \right) = 0 \ .
\ee
This shows that $Z$ depends only on $z$ and is therefore a complex analytic
function. It follows that the remaining Killing vector equation
(\ref{eq:kveq}) reduces to a form analogous to the integrability condition for
the second rank Killing tensors (\ref{eq:intcond2}). It can also be written
in the form
\be
     K_{(1;2)} = G^{-1} \Re\{ G Z'(z) + G_{,z} Z(z) \}
               = G^{-1} \Re\{(GZ)_{,z}\} = 0 \ .
\ee
Any Killing vector gives rise to a second rank Killing tensor given by
\be  K_{IJ} = K_I K_J + c h_{IJ} \ ,
\ee
where $c$ is an arbitrary (real) constant. The conformal Killing tensor
components then become
\be\eqalign{
              P_{11} &= K_{11} = (K_1)^2 = G \bar Z^2 \ ,\cr
              P_{22} &= K_{22} = (K_2)^2 = G      Z^2 \ ,\cr
}\ee
Referring to (\ref{eq:kt}) it follows that the conformal Killing tensor is
determined by the analytic function
\be  S = Z^2 \ .
\ee

As in the second rank case, invariance at arbitrary energy involves a
further restriction coming from the requirement that (\ref{eq:kveq})
should be invariant with respect to energy redefinitions. This means that we
must impose the condition
\be  \Re[Z'(z)] = 0 \ ,
\ee
leading to the simplified Killing vector equation
\be  G_{,z} Z + G_{,\bar z} \bar Z = 0 \ .
\ee
where
\be  Z(z) = U(z) := ibz + \delta \ ,
\ee
and $b$ and $\delta$ are real and complex constants respectively. It follows
that the corresponding second rank conformal Killing tensor is determined by
\be S(z) = [U(z)]^2 = -b^2 z^2 + 2ib\delta z + \delta^2 \ .
\ee
Thus $S(z)$ is an even square for a reducible Killing tensor so the condition
$\Delta^2 = \beta^2/4 - a\gamma = 0$ is always satisfied.

\section{Transforming Hamiltonian symmetries between different time gauges}
\label{sec:timetrans}

In this section we address the problem of how an invariant is affected when
we transform from one time gauge to another. Suppose we are given a
Hamiltonian $H$ with a second invariant $I$ so that $\{I,H\} =0$. In another
time gauge the Hamiltonian is given by $\HNscr = \Nscr\Hscr = \Nscr (H-E)$
where $\Nscr = dt/d\tilde t$. Of particular interest for the purposes of this
paper is the transformation between the physical and Jacobi time gauges. In
that case, going from the physical time gauge to the Jacobi time gauge, $dt =
\Nscr dt_J$, we have $\Nscr = |2V|^{-1}$. Since the Poisson bracket
\be\label{eq:comm}
     \{I,\HNscr\} = \Hscr \{I,\Nscr\} \ ,
\ee
is in general non-vanishing off the zero energy surface $\Hscr =0$, the
original invariant does not have a vanishing Poisson bracket with the
Hamiltonian in the new time gauge. Borrowing usage from Dirac's theory of
constrained Hamiltonians we may say that the invariant is only weakly
conserved in the new time gauge. For a linear invariant, e.g.\ the
Killing vector case, one can always choose variables such that $I = p_{\tilde
x}$ and $\tilde x$ is a cyclic variable in the Hamiltonian. Then for the
Jacobi time gauge, $\Nscr$ depends only on the remaining variable and hence
$\{I,\HNscr\} =0$ implying that $I$ is strongly conserved in the new time
gauge. Suppose now that we have a 2-dimensional system with a second
invariant given in the physical time gauge. Then, the system is integrable
and we can express the lapse explicitly as a function $\Nscr = f(t)$ of the
physical time $t$. Integrating the relation $d\tilde t = dt/f(t)$ then gives
the new time as an explicit function of the old time at least up to a
quadrature. In principle, the solutions can consequently always be
transformed to another time gauge. One therefore expects that the
integrability properties of a dynamical system are independent of the time
gauge. In order to exploit fully the geometrical formulation of mechanics it
is desirable to find a corresponding invariant which is strongly conserved in
the Jacobi time gauge. As we shall see this is actually possible at least in
the cases considered in this paper.

To find the invariant $\IN$ in the new time gauge it is convenient to
use an ansatz of the form
\be\label{eq:invrelation}
     \IN=I+R\HNscr \ .
\ee
We look for a condition on $R$ which guarantees that $\IN$ is a constant of
the motion for $\HNscr$. To that end we form the Poisson bracket
\be
     \{\IN, \HNscr\} = \Nscr\HNscr\Bigl[\{R, \Hscr\} - \{I,\Nscr^{-1}\}
                              - \HNscr\{R,\Nscr^{-1}\}\Bigr].
\ee
Requiring this expression to vanish off the zero energy surface yields the
condition
\be\label{eq:Rcond}
    \{R, \Hscr\} = \{I,\Nscr^{-1}\} + \HNscr\{R,\Nscr^{-1}\}.
\ee
This is a necessary and sufficient condition for $\IN$ to be a constant of the
motion for $\HNscr$. In general a solution to this partial
differential equation would be difficult to find. For our purposes,
however, it turns out that we can actually find a solution by a simple
procedure which we now outline. Suppose now that $I$ is a quadratic
invariant. Then if $\Nscr$ is a function on the configuration space, the
bracket $\{I,\Nscr^{-1}\}$ is a linear function of the momenta. Suppose we
find a function $R$ on the configuration space which satisfies $\{R, \Hscr\} =
\{I,\Nscr^{-1}\}$. Then since $\HNscr\{R,\Nscr^{-1}\} =0$ we have a
solution of (\ref{eq:Rcond}). In this way the procedure to find the
invariant in the new time gauge is reduced to calculating the Poisson
bracket $\{I,\Nscr^{-1}\}$ and finding a function whose time derivative
coincides with that bracket. Although we do not know under which conditions
this procedure works it does work in the cases considered in this paper.
To summarize we first calculate the function $\{I,\Nscr^{-1}\}$
and check whether it can be expressed as a total time derivative of some
function $R$. If in addition $\{R,\Nscr^{-1}\}=0$ then $R$ is the required
function which satisfies (\ref{eq:Rcond}).

\section{The relation between the quadratic invariants in the physical and
         Jacobi time gauges}

We wish to see how the invariant (\ref{eq:invreal}) appears in the physical
time gauge in terms of the conformal function $S(z)$ and the trace $K$. To
that end we use the procedure outlined in section \ref{sec:timetrans} going
``backwards" from the Jacobi time gauge to the physical time gauge.
Referring to section \ref{sec:geom} the starting point is now the Jacobi
Hamiltonian $H = H_J = \frac14 G^{-1} (p_x{}^2 + p_y{}^2)$ considered at
the energy value $1/2$ so $\Hscr = H_J - 1/2$. The transformation from the
Jacobi time to the physical time is then given by $dt_J = \Nscr dt$ where
$\Nscr = 2G$. According to (\ref{eq:invrelation}) we write the physical time
invariant as $I = I_J + R\Nscr \Hscr$. Using the prescription in
section \ref{sec:timetrans}, $R$ should satisfy the equation $\{I_J,
\Nscr^{-1}
\} = \{R, H_J\}$. Calculating the left hand side $\{ I_J, \Nscr^{-1} \}$ with
$I_J$ given by the expression (\ref{eq:invreal}) gives
\be\label{eq:IJNpoisson}
     \{I_J, \Nscr^{-1} \} = G^{-3}(SGG_{,z} + \frac12 K G_{,\bar z}) p_z
               + G^{-3}(\bar S GG_{,\bar z} + \frac12 K G_{,z}) p_{\bar z} \ .
\ee
Assuming that $R$ is a function on the configuration space the right hand
side is given by
\be\label{eq:rhs}
     \{R, H_J \} = G^{-1} (R_{,\bar z} p_z + R_{,z} p_{\bar z}) \ .
\ee
Comparing equations (\ref{eq:IJNpoisson}) and (\ref{eq:rhs}) then leads to
\be\label{eq:Req}\eqalign{
       R_{,z} &= \bar S G^{-1} G_{,\bar z} + \frac12 K G^{-2} G_{,z} \ ,\cr
       R_{,\bar z} &= S G^{-1} G_{,z} + \frac12 K G^{-2} G_{,\bar z} \ .\cr
}\ee
At this point it is convenient to introduce a function $Q = R +
(1/2)KG^{-1}$. The equations (\ref{eq:Req}) then reduce to
\be\label{eq:Qeq}\eqalign{
     Q_{,z}      &= -\frac12 \bar S_{,\bar z} \ ,\cr
     Q_{,\bar z} &= -\frac12 S_{,z}
}\ee
where we have used the trace equation (\ref{eq:traceeq}). The integrability
condition for this equation coincides with the condition for integrability
at arbitrary energy, $\Im S''(z) = 0$. Using (\ref{eq:poly}) we then find that
the solution of (\ref{eq:Qeq}) is given by
\be
     Q = - a z \bar z -\frac12 \bar\beta z - \frac12 \beta \bar z + Q_0 \ ,
\ee
where $Q_0$ is a (real) integration constant.

Collecting our results we find that the expression for the invariant in the
physical time gauge is
\be
     I = I_{conf} + \frac12 Q (p_x{}^2 + p_y{}^2) + \frac{K}2 - GQ \ ,
\ee
where
\be\label{eq:Iconf}
     I_{conf} = P^{MN} p_M p_N =
                \frac12 \Re(S) (p_x{}^2 - p_y{}^2) + \Im(S) p_x p_y \ ,
\ee
is the conformal part of the invariant. It follows that this part is the
same in both time gauges. It is also seen that $Q$ can be identified with
the trace of the physical invariant with respect to the physical metric
$\delta_{ab}$ where $a$ and $b$ are coordinate indices taking the values $x$
and $y$. The relation between the invariants provides a shortcut to
calculate the Killing tensor from knowledge of the physical invariant. We
shall outline this procedure and then illustrate with an example. Let the
physical invariant be given by an expression of the form
\be
     I = Q^{ab}(x,y) p_a p_b + f(x,y) \ ,
\ee
The physical Hamiltonian is given by
\be
     H = \frac12 \delta^{ab} p_a p_b + V(x,y) \ ,
\ee
We can read off the function $Q$ by $Q = \delta^{ab}Q_{ab}$. Taking the
conformal part and identifying with (\ref{eq:Iconf}) we obtain
\be\label{eq:Sconf}
     S = 2(P^{xx} + i P^{xy}) \ ,
\ee
where $P^{ab} = Q^{ab} - \frac12 Q \delta^{ab}$ and $P^{yy} =
-P^{xx}$. Finally we obtain the Killing tensor trace as $K = 2(f +
GQ)$ putting $G = E-V$.

Let us now illustrate the above results by an example. We take the Kepler
potential $V = -\mu/r$ where $r = \sqrt{x^2+y^2}$ with its well-known
non-trivial quadratic invariant, the Laplace-Runge-Lenz vector (see e.g.\
\cite{goldstein:mech2}) with components
\be\label{eq:LRLx}\eqalign{
     L_1 := e_x &= \frac1\mu (x p_y{}^2 - y p_x p_y) - \frac{x}{r} \ ,\cr
     L_2 := e_y &= \frac1\mu (y p_x{}^2 - x p_x p_y) - \frac{y}{r} \ .\cr
}\ee
The homogeneous and inhomogeneous parts of the invariants $L_1$ and
$L_2$ are consequently given by
\be
     Q_{(1)}^{ab} = \left( {\matrix{ {0}           & {-y/ (2\mu )} \cr
                                     {-y/ (2\mu )} &        x/ \mu \cr
                  }} \right) \ ,\qquad
    f_{(1)} = -x/r \ ,
\ee
and
\be
     Q_{(2)}^{ab} = \left( {\matrix{ {y/\mu}           & {-x/ (2\mu )} \cr
                                         {-x/ (2\mu )} &             0 \cr
                  }} \right) \ ,\qquad
    f_{(2)} = -y/r \ .
\ee
It follows that the conformal Killing tensor components are given by
\be\label{eq:confcomp12}\eqalign{
     P_{(1)}^{xx} &= -P_{(1)}^{yy} = -x/(2\mu) \ ,\qquad
     P_{(1)}^{xy}  = -y/(2\mu) \ ,\cr
     P_{(2)}^{xx} &= -P_{(2)}^{yy} = y/(2\mu) \ ,\qquad
     P_{(2)}^{xy}  = -x/(2\mu) \ ,\cr
}\ee
while the traces are given by
\be\label{eq:trace12}
     K_{(1)} = 2(E/\mu)x \ ,\qquad K_{(2)} = 2(E/\mu)y \ .
\ee

\section{Applications to integrability at fixed energy}
An important lesson to be learned from the present work is that
integrability at fixed energy and arbitrary energy (weak and strong
conservation laws) are just two aspects of the same phenomenon. In
particular a fixed energy invariant in the physical time gauge corresponds to
an arbitrary energy invariant if the system is geometrized by going to the
Jacobi time gauge. To illustrate how this works in practice we give a few
examples in section \ref{sec:degeneracy} of fixed energy invariants beginning
with the Kepler potential. Surprisingly, we find two apparently unknown linear
invariants at zero energy for the Kepler problem. It is remarkable that it is
still possible to discover new properties of such a simple and well-known
system. This is in fact a sign of the power of the geometric formulation of
dynamical systems. In section \ref{sec:otherfixedinv} we discuss how conformal
transformations can be used to generate systems which are integrable at
fixed energy.

\subsection{Degeneracy of the Laplace-Runge-Lenz vector at zero energy}
\label{sec:degeneracy}

Consider now the Kepler potential, $V = -\mu/r$, using polar coordinates
defined by $x = r\cos\phi$, $y = r\sin\phi$. In this case we have one linear
invariant at arbitrary energy, the angular momentum $p_\phi$. We are
interested in finding out whether there exists another linear invariant at
some fixed energy value. If this is the case the Jacobi metric
(\ref{eq:jacobimetric}) has two Killing vectors. However, a 2-dimensional
space with two Killing vectors must necessarily also have a third Killing
vector (\cite{ksmh:exact}, Theorem 8.15) and the geometry is then a space of
constant curvature. This can easily be determined by computing the scalar
curvature
\be
     \twoR = 2G^{-2}G_{,z\bar z} - 2G^{-3}G_{,z}G_{,\bar z}   \ ,
\ee
of the metric (\ref{eq:ds2complex}) and checking if it is constant. For the
Kepler potential we have $G = E + \mu (z\bar z)^{-1/2}$ and the scalar
curvature becomes
\be
     \twoR = -\frac{E\mu}{2[\mu + E(z\bar z)^{1/2}]^3} \ .
\ee
This shows that the Jacobi geometry has constant curvature only if $E=0$ and
then the geometry is actually flat. Of course the flatness of the Jacobi
geometry in this case also follows directly from the form of the metric since
$G$ is then a product of functions of $z$ and $\bar z$. The two extra Killing
vector fields can immediately be written down if we introduce Cartesian
coordinates,
$(X, Y)$, for the Jacobi geometry by the transformation $w = X+iY = \sqrt{2z}$
leading to the manifestly Euclidean form
\be
     ds_J{}^2 = 4\mu (d X^2 + d Y^2) \ .
\ee
The relation to the original coordinates is given by the parabolic
transformation
\be
     x = \case12 (X^2 - Y^2) \ ,\qquad  y = XY \ .
\ee
This transformation does not have a unique inverse. However, in the region
$X = \Re(w) >0$ we may select the inverse transformation to be
\be
     X = \sqrt{r+x} \ ,\qquad  Y = (\sgn y)\sqrt{r-x} \ .
\ee
It follows that the Killing vector fields are $K_{(1)} = \partial/ \partial
X$, $K_{(2)} = \partial/ \partial Y$ and $K_{(3)} = -Y \partial/ \partial X +
X \partial/ \partial Y = 2\partial/\partial\phi$. The two extra Killing vector
fields are thus the translation symmetries $K_{(1)}$ and $K_{(2)}$. The
corresponding invariants are
\be\label{eq:keplerinv}\eqalign{
     I_1 &:= p_X =  (r+x)^{1/2} p_x + (\sgn y) (r-x)^{1/2} p_y
                = (2r)^{1/2} \cos(\phi/2) p_r
                                    - (2/r)^{1/2} \sin(\phi/2) p_\phi \ ,\cr
     I_2 &:= p_Y = -(\sgn y)(r-x)^{1/2} p_x + (r+x)^{1/2} p_y
                = (2r)^{1/2} \sin(\phi/2) p_r
                                    + (2/r)^{1/2} \cos(\phi/2) p_\phi\ .\cr
}\ee
These two invariants are related by the quadratic formula $I_1{}^2 + I_2{}^2 =
8\mu H_J^0$ where $H_J^0 = -(2V)^{-1}T = \frac12$ is the Jacobi Hamiltonian
for the zero energy system. In terms of the physical Hamiltonian, the
corresponding relation is $I_1{}^2 + I_2{}^2 = 4\mu(1-H/V) = 4\mu$.
In fact here we have the key to the physical interpretation of $I_1$ and
$I_2$. To see this let us introduce an invariant $\phi_0$ at zero energy by
the relations
\be\label{eq:inv1}
     I_1 = -2\mu^{1/2} \sin(\phi_0/2) \ ,\qquad
     I_2 =  2\mu^{1/2} \cos(\phi_0/2) \ .
\ee
Now using (\ref{eq:keplerinv}) to solve for the radial momentum yields $p_r =
\sqrt{2\mu/r} \sin[(\phi-\phi_0)/2]$. Inserting this value into the
Hamiltonian constraint $H=0$ and solving for $r$ gives the familiar relation
\be
     r = \frac{p_\phi{}^2/\mu}{1+\cos(\phi-\phi_0)} \ .
\ee
This shows that $\phi_0$ is nothing but the angular integration constant.
It follows that changing the value of $I_1$ (or $I_2$) only affects the
parametrization of the orbit while leaving the orbit itself invariant. From
this point of view these invariants are gauge symmetries of the Kepler system
at zero energy.

The invariants $I_1$ and $I_2$ are in fact closely related to the components
of the Laplace-Runge-Lenz vector. Expressing those components as
\be
     e_x = e\cos\phi_0 \ ,\qquad e_y = e\sin\phi_0 \ ,
\ee
where $e$ is the eccentricity ($e=1$ at zero energy) and comparing with
(\ref{eq:inv1}) it is evident that
\be\label{eq:LRLxy}
     L_1 = -(2\mu)^{-1} I_1{}^2 + 1 \ ,\qquad
     L_2 = -(2\mu)^{-1} I_1 I_2 \ .
\ee
This can also be seen directly by calculating for example $I_1{}^2$ from the
expression given in (\ref{eq:keplerinv}) with result
\be
     I_1{}^2 = 2(r+x)H + 2\mu(1-L_1) \ ,
\ee
where we have used (\ref{eq:LRLx}). Now solving for $L_1$ at zero energy
gives again the first of the relations (\ref{eq:LRLxy}).

The relations (\ref{eq:LRLxy}) imply that the second rank Killing tensors
corresponding to the components of the Laplace-Runge-Lenz vector are
reducible at zero energy. The invariants in the Jacobi time gauge can be found
from the relations (\ref{eq:confcomp12}), (\ref{eq:trace12}),
(\ref{eq:Sconf}) and (\ref{eq:invreal}). This gives
\be\eqalign{
     J_1 &= -(2\mu)^{-1} x(p_x{}^2-p_y{}^2) - \mu^{-1} y p_x p_y \ ,\cr
     J_2 &= (2\mu)^{-1} y(p_x{}^2-p_y{}^2) - \mu^{-1} x p_x p_y \ ,\cr
}\ee
where $J_1$ and $J_2$ are the Jacobi invariants corresponding to $L_1$ and
$L_2$ respectively. From (\ref{eq:keplerinv}) we then find the relations
\be\eqalign{
     J_1 &= -(2\mu)^{-1} I_1{}^2 + 2 H_J \ ,\cr
     J_2 &= -(2\mu)^{-1} I_1 I_2 \ .\cr
}\ee
The reducibility of the Killing tensors $K_{(1)}^{MN}$ and $K_{(2)}^{MN}$
corresponding to $J_1$ and $J_2$ is therefore expressed by the formulas
\be\eqalign{
     K_{(1)}^{MN} &= -(2\mu)^{-1} K_{(1)}^M K_{(1)}^N + J^{MN} \ ,\cr
     K_{(2)}^{MN} &= -(2\mu)^{-1} K_{(1)}^{(M} K_{(2)}^{N)} \ .\cr
}\ee

We also wish to understand the commutation relations for the fixed energy
invariants in the physical time gauge. To facilitate the calculations the
physical Hamiltonian  is first expressed in terms of the Jacobi Hamiltonian by
\be
     \Hscr = T + V - E = (V-E)(1-2H_J) \ ,
\ee
where we have used $T = 2(E-V)H_J$. We can now exploit the fact that the
invariant commutes with the Jacobi Hamiltonian to obtain
\be\label{eq:fixedbracket}
     \{I, H\} = \{I, \Hscr\} = (1-2H_J) \{I, V\}
                             = \Hscr(V-E)^{-1} \{I, V\} \ .
\ee
This relation shows that we only need to compute the Poisson bracket with the
potential. It also follows that the bracket $\{I, H\}$ in general depends
linearly on $\Hscr$.

Using (\ref{eq:fixedbracket}) to calculate the brackets for the Kepler
invariants we find
\be\eqalign{
     \{I_1, H\} &= r^{-1}(r+x)^{1/2} H
                 = \sqrt2 r^{-1/2} \cos(\phi/2) H \ ,\cr
     \{I_2, H\} &= -(\sgn y)r^{-1}(r-x)^{1/2} H
                 = -\sqrt2 r^{-1/2} \sin(\phi/2) H \ .\cr
}\ee

\subsection{Some other examples of integrability at fixed energy}
\label{sec:otherfixedinv}

As discussed for example by Hietarinta \cite{hietarinta:secinv}, conformal
transformations provide links between physically different systems which are
integrable at some fixed energy. In particular if the original potential
$\tilde V$ is separable, $\tilde V=Q_1(X)+Q_2(Y)$, then the transformed
system has the potential
\be
     V = |H'(z)|^2 [Q_1(\Re(H(z)) + Q_2(\Im(H(z)) - E] \ ,
\ee
where $z=x+iy$ and $H(z)= X+iY$. One of the results of the present work is
that we have identified those conformal transformations for which the new
potential in this situation is actually integrable at arbitrary energy (see
table \ref{tab:conf}). Conversely, if the conformal transformation is not
contained in table \ref{tab:conf} then the resulting potential does not have a
linear or quadratic invariant at arbitrary energies. Hietarinta considered
conformal transformations of the forms $H(z) = z^m$ (with $m = -1, -2,
\frac12, 2$), $e^z$ and $\ln z$. Note in particular that of these $H(z) =
z^{1/2}$ and $H(z) = \ln z$ produce systems which are integrable at arbitrary
energy if the original potential is separable. In this subsection we give
some further examples of simple systems which are integrable at a fixed
energy.

Consider first the polynomial function $S(z) = i z^2$. This is the simplest
polynomial which gives a potential which is not automatically integrable at
arbitrary energy. The corresponding conformal transformation is given by $w =
H(z) = 2^{-1/2} (1-i) \ln z$ or in terms of the real variables
\be\eqalign{
      X &= \frac1{\sqrt2} (\theta + \ln r) \ ,\cr
      Y &= \frac1{\sqrt2} (\theta - \ln r) \ .\cr
}\ee
From the relation (\ref{eq:potrel}) it then follows that the potential
given by
\be
      G = r^{-2} [ A(r e^{\theta}) + B(r e^{-\theta}) ] \ ,
\ee
is integrable at zero energy for arbitrary functions $A$ and $B$. For
functions containing linear and quadratic terms the potential takes the form
\be
     G = r^{-1} (a_1 e^{\theta} + a_2 e^{-\theta} ) + 
                 a_3 e^{2\theta} + a_4 e^{-2 \theta} \ ,
\ee
where the $a_i$ ($i=1,\ldots,4$) are arbitrary constants.

As another example we take a function of the form $S(z) = z^k$ where $k \neq
0,1,2$ is a real constant. This leads to $w = H(z) = m^{-1} z^m$ where $m =
-k/2+1 \neq 0,\frac12,1$. The conformal transformation can then be written
\be\eqalign{
     X &= m^{-1} r^m \cos(m\theta) \ ,\cr
     Y &= m^{-1} r^m \sin(m\theta) \ .\cr
}\ee
The corresponding potential is
\be
     G = r^{-k} [ A(X) + B(Y) ] \ .
\ee
Choosing for example $A(X) = a_1 m^s X^s$ and $B(Y) = a_2 m^s Y^s$ we have
\be
     G = r^{-2+m(s+2)} [a_1 \cos^s(m\theta) + a_2 \sin^s(m\theta)] \ .
\ee
Specializing to the case $m=2$ while keeping $s$ arbitrary and using
$\cos(2\theta) = (x^2-y^2) r^{-2}$, $\sin(2\theta) = 2xyr^{-2}$ yields finally
the potential
\be
     G = r^2 [a_1 (x^2-y^2)^s + a_2 2^s x^s y^s ] \ ,
\ee
which is therefore integrable at zero energy.

\section{Concluding remarks}

It was shown in section \ref{sec:posdef} that conformal transformations given
by analytic functions $H(z)$ for which the condition $\Im\{S''(z)\} = 0$ with
$S(z) = [H'(z)]^{-2}$ is satisfied give rise to the classical potentials
which admit quadratic (or linear) second invariants at arbitrary energies.
Conformal transformations which do not satisfy that condition give potentials
which admit quadratic second invariants only at a fixed energy. Our approach
unifies the description of quadratic invariants at arbitrary and fixed
energies. In particular the integrability condition (\ref{eq:intcond2}) is
valid for both types of invariants. It reduces to Darboux's classical
condition for arbitrary energy invariants when $\Im\{S''(z)\} = 0$. Whether a
unification can also be achieved for third degree invariants or higher
remains an open problem.

An intriguing aspect of the integrability condition (\ref{eq:intcond2}) is
the possibility that for a given potential function $G = E-V$ there could
exist a family of solutions $S(z,E)$ with a continuous dependence on the
energy. This would lead to new families of integrable potentials with energy
dependent quadratic invariants. At this point we cannot exclude the existence
of such solutions of the integrability condition. A related result was given
by Hietarinta \cite{hietarinta:secinv} who showed that the potential $x/y$ is
in fact integrable by energy dependent invariants which are certain
transcendental functions of the momenta.

For a given potential $V$ the integrability condition (\ref{eq:intcond2})
with $G = E-V$ can be used to determine energy values for which there exists
an invariant of at most second degree. This involves solving a linear
differential equation of the second order. For linear invariants it is
sufficient to solve the linear equation (\ref{eq:kveq}). We consider this
possibility to test for linear and quadratic integrability at fixed energy to
be an important application of the geometric approach to Hamiltonian
dynamics. Another approach is to look for conditions involving curvature
invariants such as the technique used in section \ref{sec:degeneracy} to
find cases with additional invariants. In fact, conditions involving
curvature invariants for (1+1)-dimensional models have recently been found
(using two different approaches) for potentials which do not require the
manifest linear invariant present in the Kepler problem
\cite{gr:ministar,ubm:classi}. There are indications that at least the
approach of \cite{gr:ministar} can be generalized to incorporate quadratic
invariants as well. It is of considerable interest to develop these
techniques and use them to look for fixed energy invariants of physically
interesting models such as the H\'enon-Heiles potential and others.

\end{document}